October 20, 2026

# Estimand frameworks for causal inference in clinical research and regulatory decision making

Miguel A. Hernán,[1,2] Mats J. Stensrud,[3] Xabier García-Albéniz,[4]
Issa J. Dahabreh,[1,2,5] Jonathan A.C. Sterne[6,7,8]

1. CAUSALab, Department of Epidemiology, Harvard T.H. Chan School of Public Health, Boston, MA, USA
2. Department of Biostatistics, Harvard T.H. Chan School of Public Health, Boston, MA, USA
3. Institute of Mathematics, Ecole Polytechnique Fédérale de Lausanne, Lausanne, Switzerland.
4. Adigens Health, Dublin, Ireland
5. Richard A. and Susan F. Smith Institute, Beth Israel Deaconess Medical Center, Boston MA
6. Electronic Health Records Group, Population Health Sciences, Bristol Medical School, University of Bristol, Bristol, United Kingdom
7. Health Data Research UK, United Kingdom
8. NIHR Biomedical Research Centre Bristol, United Kingdom

**Word count:** 3938

**Running title:** Estimand frameworks for causal inference

**Correspondence:** Miguel Hernán, Department of Epidemiology, Harvard T.H. Chan School of Public Health, Boston, MA 02115; +1 617 432 0101; miguel_hernan@post.harvard.edu

**Funding sources:** This work was supported by National Institutes of Health grants R37 AI102634 and R01HL136708, and by Patient-Centered Outcomes Research Institute (PCORI) award ME-2021C2-22365.

**Acknowledgments:** Miguel Hernán is advisor to ProPublica, consultant and shareholder of Adigens Health, member of ADIALab's Advisory Board, and editor at Annals of Internal Medicine. His research is funded by the U.S. National Institutes of Health, U.S. Office of Naval Research, Novo, and PCORI.

**Abstract**

Causal inference for clinical and regulatory decision making requires a precise definition of the causal estimand, that is, what we want to estimate. Three frameworks are commonly used to describe causal estimands: PICO and its extensions, the target trial framework, and the E9(R1) Addendum of the International Council for Harmonisation (ICH). Because these frameworks were developed by different scientific communities, they operate at different levels of detail, emphasize different aspects of the causal estimand, and use different terminology for closely related concepts. Here we compare the three frameworks and translate across their terminologies.

PICO is a compact framework that helps jumpstart a conversation about the causal question, but it does not encode sufficient information to fully characterize a causal estimand. The target trial framework characterizes the causal estimand precisely by specifying the components of the protocol of the randomized trial that would answer the causal question: eligibility criteria, treatment strategies, assignment, outcome, start and end of follow-up, and causal contrast. E9(R1) starts from intercurrent events and lists strategies to address them, some of which describe estimators rather than estimands, or imply estimands that are not relevant to decision makers or cannot be estimated under reasonable assumptions. We argue that the target trial framework subsumes the other two, and we provide a map of the correspondences across the three frameworks.

## Introduction

In Douglas Adams's book *Hitchhiker's Guide to the Galaxy*,[1] aliens embark on a quest for the answer to "the ultimate question of life, the universe, and everything." Eventually, they obtain it: the answer was 42. Unfortunately, the question was too vague. A similar situation has been common in medical research on Earth. Investigators sometimes ask vague questions about "the association between treatment and outcome" and use sophisticated tools (e.g., Cox models with time-varying variables) that produce precise answers (e.g., "the hazard ratio is 0.73"). But, with a poorly specified question, it is often unclear how the answer can inform clinical or regulatory decisions.

Addressing this requires a precise definition of the *estimand* of interest. The estimand is what we want to estimate. In clinical research, a causal estimand is typically the causal effect of a health care intervention, and is estimated via randomized trials or observational (non-interventional) studies.

Three frameworks commonly used to describe the causal estimand of such studies are PICO,[2,3] the target trial framework,[4,5] and the E9 (R1) Addendum[6] of the International Council for Harmonisation (ICH) of Technical Requirements for Registration of Pharmaceuticals for Human Use. Each helps articulate a causal question, but they were developed by different scientific communities. It is therefore unsurprising that the frameworks operate at different levels of detail, emphasize different aspects of the causal estimand, and often use different terminology for closely related concepts.

Here we compare these three estimand frameworks for causal inference, translate across their terminologies, and highlight their commonalities and peculiarities. We argue that, for decision makers who ask causal questions, the target trial framework subsumes the other two by offering the highest level of detail and giving the clearest guidance for data analysis.

## PICO and its extensions: an abridged framework for causal estimands

The first step for causal inference is specifying the components of an unambiguous estimand, prominently the 1) population of individuals to which the causal effect applies, 2) actions compared, and 3) outcome studied.

For example, consider the causal effect of receiving semaglutide compared with an older therapy (the actions) on the risk of cardiovascular disease (the outcome) among individuals with diabetes (the population). The PICO acronym encodes these three components: P (population); I (intervention) and C (comparator) for the actions, and O (outcome). PICO was originally developed to help clinicians map a patient's scenario into a structured causal question, which could then be used to effectively query medical research databases .[2,7] PICO was subsequently widely adopted by organizations that generate or evaluate systematic reviews (e.g., Cochrane,[8] the PRISMA Statement, the World Health Organization Guideline Review Committee) or that formulate clinical questions (e.g., AHRQ, CDC, GRADE,[9] NICE).

PICO's simple structure, is often misread as a license to oversimplify the description of the actions I and C. For example, when studying diabetes therapies, specification of a treatment regime with multiple components may be tersely summarized under PICO as "Semaglutide therapy" which, as described below, is an incomplete characterization of a treatment strategy. These broad descriptions can be useful when attempting to apply a single label across diverse studies (as is often the case in systematic reviews); less so when attempting to plan, conduct, or interpret an individual study.

Because more components are needed to describe the estimand, PICO has been extended in various ways. For example, in the PICOTS extension, T (time, timeframe, or timing) refers to the duration of the intervention, the duration of the follow-up, or the time when the outcome is measured, and S (setting) refines the population by specifying where the study is conducted (e.g., primary care, hospital, nursing home). In a proposed PICOTT extension, a second T stands for type of study design, and in a proposed PICOTSI extension,[10] a second I for intercurrent events.

Because continuing to add letters to PICO would be unwieldy, a more richly structured framework that can help investigators specify the causal estimand is called for.

**Target trial: an unabridged framework for causal estimands**

For decades, clinical researchers have specified components of the causal estimand of a randomized trial by writing the trial's protocol, including eligibility criteria, treatment strategies, assignment to the strategies, outcome, start/end of follow-up, and causal contrast (traditionally, intention-to-treat). Protocols also describe procedures to generate and analyze the data, but here we consider only the components that define the causal estimand.

The concept of the trial protocol is also relevant for observational studies that aim to quantify causal effects of medical interventions. These studies can be viewed as an attempt to emulate a hypothetical randomized trial—the target trial.[4] Therefore, a natural way to articulate the causal estimand of observational studies is to specify the protocol of the target trial that they attempt to emulate. A standardized presentation of the target trial protocol components[5] has been adopted or considered by health technology assessment agencies,[11,12] regulators,[13] the Cochrane Collaboration,[14] and others.

The Table lists the components of the target trial protocol that define the causal estimand along with their PICO counterparts. Unlike typical applications of PICO, the target trial protocol specifies the treatment strategies in detail. For example, the definition of a treatment strategy may range from "Start semaglutide therapy at any dose" to "Start semaglutide at dose *d* with potential dose adjustments [specified in the protocol] every 3 months based on clinical factors [specified in the protocol]; discontinue semaglutide or switch to an alternative therapy in the presence of serious adverse events or lack of effectiveness [also defined in the protocol])". A detailed description of the treatment strategies is necessary to specify the causal estimand and to identify the data, assumptions, and analytic approaches required to validly estimate it.

The various aspects of the causal estimand related to time in PICOTS are specified as part of each target trial component: "Treatment strategies" encompasses the timing of the intervention, "Start and end of follow-up" the timing of follow-up, and "Outcome" the timing of outcome ascertainment.

*Causal contrasts*

Unlike PICO and its extensions, the target trial framework requires specification of the causal contrast, which is a critical component of the causal estimand in both randomized trials and observational emulations. The causal contrasts used for clinical and regulatory decision making are, almost without exception, contrasts of outcome distributions under different treatment strategies. Because they are quantified as ratios or differences of outcome functionals (e.g., a risk ratio), they are population effects rather than individual effects.

Two causal contrasts are the effect of assignment and the per-protocol effect. To describe them, consider a randomized trial in which individuals are assigned to either: 1) the semaglutide strategy described in the previous section; or 2) to a similar strategy that replaces semaglutide with another drug. The outcome is the 5-year risk of cardiovascular disease.

*The effect of assignment (intention-to-treat effect)*

Consider comparing outcome risks between groups assigned to strategy 1) or 2), regardless of treatment received. The causal contrast implied by this comparison is *the effect of assignment* to a treatment strategy, not that of actually receiving a treatment strategy.[15]

In randomized trials, the effect of assignment is often referred to as the intention-to-treat effect. In the absence of losses to follow-up or other missing data, this effect can be validly estimated under essentially no assumptions. Many randomized trials are designed, conducted, and analyzed to quantify, exclusively or primarily, a causal estimand based on the intention-to-treat effect.

In observational emulations of target trials, an analog of assignment may be prescription. For example, eligible individuals could be assigned to (classified into) treatment strategies 1) and 2) based on whether they received a prescription of semaglutide or the other therapy, respectively. The effect of being prescribed the treatment, regardless of whether it is received or continued, may be viewed as analogous to the effect of assignment in the corresponding target trial.

The effect of assignment is often said to be "biased towards the null" because its magnitude decreases with the proportion of individuals who do not adhere to their assigned treatment strategy. However, even in the presence of substantial nonadherence, the effect of the assignment is not "biased" — it quantifies the effect of assignment. Concern about dependence of the effect of assignment on adherence arises because decision makers — clinicians, patients, regulators — are often interested in another causal contrast.

*The per-protocol effect*

The contrast of outcome risks if all individuals had received their assigned treatment strategies as defined in the trial's protocol is the per-protocol effect.[16,17] In randomized trials, the per-protocol effect is of interest because the investigators designed the trial with that particular protocol. Had they been interested in a different causal contrast, they would have formulated a different protocol. When the effect of assignment is said to be "biased towards the null", it is implicitly with respect to the per-protocol effect.[15]

Even when the per-protocol effect is relevant for decision makers, the effect of assignment is often prioritized in randomized trials because estimating per-protocol effects requires additional assumptions that are not guaranteed by the trial design (typically, that confounding can be measured and adjusted for) and that cannot be verified using the trial data alone. A common misconception is that the per-protocol effect is the effect *among those who adhered* to the protocol; rather, it is the effect *if everybody had adhered* to the protocol. Therefore, commonly used naïve "per-protocol analyses" that are restricted to adherers (the so-called "per-protocol population"), without adjustment for prognostic factors, are not generally valid approaches to estimate per-protocol effects and may be biased.[16]

The same logic applies to observational emulations, where the per-protocol effect refers to the target trial protocol. It is of interest because otherwise the investigators would have specified a different protocol.

Characterizing the per-protocol effect requires a specification of the treatment strategies that is sufficiently detailed to identify whether individuals adhered to their assigned strategy over time. For example, consider the semaglutide strategy specified above. If a trial participant assigned to this strategy discontinued treatment because he did not like his doctor, he deviated from the treatment strategy (did not adhere). If another participant discontinued treatment because she experienced serious adverse events, she did not deviate from the treatment strategy (adhered). Discontinuation of treatment is nonadherence when it occurs for reasons outside the treatment strategy, but not when it is consistent with the treatment strategy.

This distinction is necessary both to characterize the per-protocol effect (a component of the estimand) and to guide the data analysis (the estimator). For example, one approach to deal with nonadherence in the analysis of randomized trials is censoring individuals when they become non-adherent and then using inverse probability weighting to adjust for time-fixed and time-varying prognostic factors associated with adherence. Under this approach, individuals who stop treatment because of personal preference are censored; those who stop because of serious adverse events are not. This is different from so-called “as-treated” analyses, which incorrectly censor individuals when they stop treatment (e.g., because of serious adverse events) rather than when they stop adhering to the protocol. Such analyses also usually lack confounding adjustment.

The same rationale applies to intervention components other than treatment discontinuation. For example, the data analysis only needs to consider initiation of a concomitant therapy if the treatment strategy of interest does not allow such initiation. Without a complete definition of adherence to the treatment strategies specified in the protocol, the per-protocol effect cannot be defined and a congruent data analysis cannot be formulated.

*Competing events*

Besides distinguishing between "assignment" or "per-protocol" effects, the definition of the causal estimand also requires specification of how competing events (e.g., death from non-cardiovascular causes when the outcome is cardiovascular disease) are incorporated into the causal contrast. Regardless of whether the causal contrast is the effect of assignment or the per-protocol effect, the causal contrast must also incorporate the competing event. Some possibilities are defining the causal contrast as: a total effect (the effect of treatment on the outcome when the outcome is defined as absent for people who experience the competing event); a controlled direct effect (the effect of treatment on the outcome if, somehow, the competing event could be prevented for everybody); or a separable direct effect (the effect of treatment on the outcome through mechanisms that do not affect the competing event). These causal estimands requires different assumptions and data analysis methods.[18,19] Importantly, competing events are not adjusted for (as it is sometimes said); rather, they are incorporated into the causal question.

**ICH E9(R1)**

The E9 guideline for clinical trials, issued by ICH in 1998,[20] considered no alternatives to the intention-to-treat effect. However, other causal estimands were often pursued when, for example, adherence to the trial protocol was imperfect. ICH revised E9 in an attempt to reduce confusion and harmonize practice by clarifying the choice of the causal estimand. The resulting ICH E9 (R1) Addendum — E9(R1) for brevity — was released in 2019.[6] Shortly thereafter, it was implemented by the European Medicines Agency, the U.S. Food and Drug Administration, the China National Medical Products Administration, and other regulators. Several papers[21,22] and books[23,24] describe its implementation.

E9(R1) defines four attributes of an estimand—population, treatment conditions, variable, and population-level summary. E9(R1)'s "Population" corresponds to the P in PICO and the eligibility criteria in the target trial framework; "Treatment conditions" approximately corresponds to the I and C in PICO and the treatment strategies in the target trial; "Variable" corresponds to the O in PICO and the outcome in the target trial; and "Population-level summary" is part of the causal contrast in the target trial.

E9(R1) also describes five possible strategies to address 'intercurrent events', which are defined as "events occurring after treatment initiation that affect either the interpretation or the existence of the measurements associated with the clinical question of interest". (Some interpretations of E9(R1) define intercurrent events as a fifth attribute of the estimand, but that does not seem to be the intent of the original guideline.) When mapped to terminology often used in the causal inference literature, intercurrent events can be roughly classified as Type 1 (nonadherence to the treatment strategies [as defined in the target trial framework]) or Type 2 (competing and truncating events, which are events that make it impossible for the outcome to occur). Missing data, including loss to follow-up, is not considered an intercurrent event by E9(R1).

Before we review the five "strategies to address intercurrent events", we relate "intercurrent event" to more well-established causal inference concepts.

Discussing nonadherence (Type 1 intercurrent events) requires detailed specification of the protocol's treatment strategies. However, the E9(R1) guideline considers intercurrent events (e.g., treatment discontinuation) *before* specifying the treatment strategies (e.g., take treatment and discontinue it if serious adverse effects arise) and outcomes that define the estimand.[25] This ordering has been a source of misunderstanding. Similar issues arise when considering competing events (Type 2 intercurrent events). For example, death from non-cardiovascular causes is a competing event if the protocol-specified outcome is cardiovascular disease, but not if it is the composite event "cardiovascular disease or death, whichever happens first." Under E9(R1), death from non-cardiovascular diseases is initially considered a (Type 2) intercurrent event, regardless of the outcome specified in the target trial protocol.

Under E9(R1), investigators list the intercurrent events, then implement an approach ("strategy") that addresses them. In contrast, under the target trial framework (and, more generally, in the causal inference literature[26,27]), investigators start by articulating their causal question (specifying the components of the target trial protocol that define the causal estimand). This is required to characterize nonadherence and competing events, that is, the post-baseline intercurrent events that need to be addressed.

While users of both frameworks may end up at the same causal estimand for the same reasons, the different framing is a source of confusion. For example, some authors[21,28-30] have mistakenly proposed that the target trial framework needs to be supplemented with “strategies to address intercurrent events”.

**Mapping E9(R1) to the target trial framework**

We now review the five “strategies to address intercurrent events” in E9(R1) and their relation to the causal estimands in the target trial framework.

*Treatment policy: Ignore nonadherence (Type 1 intercurrent events) and compare individuals according to their assignment groups.* This corresponds to the effect of assignment under the target trial framework.

*Principal stratum: Eliminate competing events (Type 2 intercurrent events) by restricting the estimates to the subgroup of individuals who would not die under any of the treatment strategies under study—a principal stratum. The same approach can be followed to eliminate nonadherence (Type 1 intercurrent events) by finding the principal stratum of individuals who would adhere regardless of treatment assignment.* Because individuals in principal strata cannot be identified in the real world, these contrasts are of no direct relevance for decision makers interested in implementing the interventions. Furthermore, inference about principal stratum estimands generally requires strong assumptions that are unverifiable even in perfectly executed randomized trials.[31-35] The target trial framework privileges estimands based on populations that can be characterized in terms of observable eligibility criteria at the time when the intervention strategies are implemented.

*Composite variable: Eliminate Type 1 or Type 2 intercurrent events by redefining the outcome variable to be a composite of the outcome of interest plus the intercurrent event.*

This approach corresponds to defining the outcome as a composite event in the target trial framework.

*Hypothetical: Suppose intercurrent events did not happen*. This encompasses consideration of different types of treatment strategies. Some of these relate to interventions that can never occur in the real world (e.g., continue treatment in the presence of serious adverse events, withhold rescue medication) whose effects therefore cannot be validly estimated under reasonable assumptions.

Other treatment strategies relate to interventions that occur in the real world (e.g., follow the protocol) and whose effects can be estimated under appropriate adjustment for time-varying prognostic factors both in randomized trials and observational emulations. These are the treatment strategies privileged by the target trial framework: the per-protocol effect is a contrast of such treatment strategies.

*While on treatment: Use only outcome data that occur before the intercurrent events*. This describes an estimator (what to do with the data), not an estimand. For Type 1 intercurrent events, censoring at nonadherence can be reverse engineered under the target trial framework to the unrealistic treatment strategy "take treatment continuously, regardless of adverse effects and lack of effectiveness" estimated under the implausible assumption that those who do and do not discontinue treatment are expected to have the same outcomes. For outcomes with repeated measures, censoring makes it hard to define an estimand because the outcome is ascertained at different times for different people. A valid data analysis (estimator) usually needs to include appropriate adjustment for the time-varying prognostic factors that affect treatment discontinuation, which is generally impossible when all or nearly all individuals who experience adverse effects or poor effectiveness discontinue treatment.

For Type 2 intercurrent events such as non-cardiovascular death, the approach is redefined as "while alive" which, under the target trial framework, can be reverse engineered to the effect of treatment if deaths could be prevented in the study population.

Because death cannot be prevented in the real world, this controlled direct effect is often not of practical interest, and therefore not a good fit for the target trial framework.

See the Table for a side-by-side correspondence between the components of the target trial protocol and E9(R1)'s four attributes and five "strategies to address intercurrent events."

**Discussion**

In the Hitchhiker's Guide to the Galaxy, aliens received the uninterpretable answer "42" after seven and a half million years' computation by a super AI. The aliens, who could not use that answer to support their decision making, then asked the AI for the question that produced the answer "42". The AI offered to design a computer that could generate the question, but said this would take 10 million additional years.

Research for clinical and regulatory decision making is in a better position. We have three estimand frameworks that help us articulate causal questions before we devote substantial resources to generating answers via randomized trials or observational emulations of target trials.

PICO (and its extensions) is a compact, simple-to-use framework. Though it does not encode sufficient information to fully characterize causal estimands, it often suffices as a first pass to jumpstart a conversation about the causal question. The target trial framework precisely characterizes the causal estimand by expanding and restructuring the components of PICOTS. The E9(R1) guideline considers intercurrent events (nonadherence, competing and truncating events) as potential problems before the components of the causal estimand are specified, and lists "strategies", which imply causal estimands, to deal with those problems.

The target trial framework specifies the components of the estimand first, then the identifying assumptions and an estimator valid under them. E9(R1) mixes up estimand and estimator because some "strategies to address intercurrent events" are defined in terms of how the data will be used, rather than as components of the causal estimand: The *While on treatment* strategy declares that data after the intercurrent event will not be used, but the implied causal contrasts may not be relevant to decision makers or may be impossible to estimate under reasonable assumptions. For this and the *Principal stratum* strategy, the implied causal estimands correspond to impossible target trials. On the other hand, E9(R1) says too little about the estimators for other "strategies to address intercurrent events", which is dangerous because some investigators might declare, for example, that they are using a "hypothetical strategy" without realizing that such approach generally requires adjustment for time-varying confounders.

Some aspects of E9(R1) are rooted in the context of regulatory decision making in which it was born. Because regulators must make decisions about drugs and other products, their ideal data would arise from a randomized trial in which all individuals take only the product of interest during the entire follow-up or until they have the outcome of interest. Deviation from that idealized setting is naturally viewed as a problem (intercurrent event). However, a causal estimand involving the strategy "take treatment continuously without receiving a rescue medication" is 1) not relevant for decision making in the real world and 2) generally impossible to estimate consistently under reasonable assumptions.

In the real world, treatment decisions must be adapted over time to the patient's clinical situation. Nobody is forced to take treatment forever regardless of serious adverse events, or deprived of receiving an available rescue medication. That is, regulatory decisions are framed in terms of "treatments", but reasonable estimands are typically defined in terms of realistic "treatment strategies". Progress in regulatory decision making requires that treatment strategies are given more prominence. The final version of E9(R1) moved in this direction when it incorporated the attribute "Treatment" after its importance was pointed out in response to an early draft of the guideline.[36]

In practice, the distinction between starting from the components of the causal estimand (target trial framework) and starting from intercurrent events (E9(R1)) is often blurred because, in observational emulations of target trial, the process of estimand construction depends intimately on the available data.[5] Specification of the target trial protocol is an iterative process in which the initial causal question is adapted to the data constraints. For example, in the presence of competing events, investigators interested in a causal contrast characterized by a separable direct effect may settle on a total effect or may change the definition of the outcome to avoid competing events if the data required to estimate separable effects is not available.

In conclusion, these three estimand frameworks—PICO, target trial, E9(R1)—help investigators ask better causal questions for clinical and regulatory decision making. Each can be applied to randomized trials and observational studies. The target trial framework subsumes the other two and clarifies when E9(R1) leads to unrealistic causal estimands. The Table provides a map across the frameworks.

**Table.** Correspondence between three estimand frameworks: PICOTS, target trial, and ICH E9(R1) Addendum

| **PICOTS** | **Target Trial** | **ICH E9 (R1) Addendum** |
|---|---|---|
| P, S | Eligibility criteria | Attribute: Population<br>Possible strategy: *Principal stratum* |
| I, C, T | Treatment strategies | Attribute: Treatment conditions<br>Possible strategy: *While on treatment* (generally not a well-defined estimand) |
| | Assignment | |
| O, T | Outcome | Attribute: Variable<br>Possible strategy: *Composite variable* |
| T | Start and end of follow-up | |
| | Causal contrast, including incorporation of competing events | Attribute: Population-level summary<br>Possible strategies: *Treatment policy*, *Hypothetical, While alive* |